\documentstyle[prc,aps]{revtex}
\begin{document}

\draft

\title{Direct Neutron Capture for Magic--Shell Nuclei}

\author{E. Krausmann, W. Balogh, H. Oberhummer}
\address{Institut f\"ur Kernphysik, Technische Universit\"at Wien, Wiedner
Hauptstrasse
8--10, 1040 Vienna, Austria}
\author{T. Rauscher}
\address{Institut f\"ur Kernchemie, Universit\"at Mainz,
Fritz-Strassmann--Weg 2,
55099 Mainz, Germany\\
and\\
Institut f\"ur theoretische Physik, Universit\"at Basel,
Klingelbergstr.\ 82, 4056 Basel, Switzerland}
\author{K.-L. Kratz, W. Ziegert}
\address{Institut f\"ur Kernchemie, Universit\"at Mainz,
Fritz-Strassmann--Weg 2, 55099 Mainz, Germany}

\maketitle

\begin{abstract}
In neutron capture for magic--shell nuclei the direct reaction mechanism can be
important and may even dominate. As an example we investigated the reaction
$^{48}$Ca(n,$\gamma)^{49}$Ca for projectile energies below 250\,keV in a direct
capture model using the folding procedure for optical and bound state
 potentials. The obtained theoretical cross sections are in
 agreement with the experimental data showing the dominance of
the direct reaction mechanism in this case. The above method
was also used to calculate the cross section for $^{50}$Ca(n,$\gamma)^{51}$Ca.
\end{abstract}


\section{Introduction}
Attempts to understand low- and medium--density ($n_{\rm n} \leq 6 \cdot
10^{16}$\, n/cm$^3$) neutron--capture nucleosynthesis processes require
temperature--dependent Maxwellian--averaged capture cross sections.
For radioactive isotopes, where these
cross sections cannot be measured they are usually calculated within global
Hauser--Feshbach models. Whereas this statistical approach appears to be
applicable for nuclides in between neutron shells exhibiting sufficiently
high level densities, it certainly will fail for neutron--magic isotopes
with only a few widely spaced resonances. For such cases apart from
compound--nucleus (CN) capture, direct (DI) reaction processes are important.

As an example we investigated the reaction $^{48}$Ca(n,$\gamma)^{49}$Ca to
test our method, where,
in principle, all necessary information is already available from
various experiments \cite{a,b,c,d,e,f,g,h,i}. We also calculated the reaction
$^{50}$Ca(n,$\gamma)^{51}$Ca, which may be of importance in connection with an
interpretation of the $^{50}$Ti overabundance observed in FUN inclusions of the
Allende
meteorite (see e.g. \cite{j,k}).

In the CN mechanism the projectile merges with the target nucleus and excites
many degrees of freedom of the CN. The excitation proceeds via a multistep
process and has a reaction time typically of the order $10^{-14}$\,s to
$10^{-20}$\,s. After this time the CN decays into various exit channels.
In the DI process the projectile excites
only a few degrees of freedom (e.g. single--particle or collective). The
excitation proceeds in one single step and has a characteristic time scale of
$10^{-21}$\,s to $10^{-22}$\,s. This corresponds to the time it takes the
projectile to pass through the target nucleus; this time is much shorter than
the reaction time of the CN processes.

For thermonuclear (in the keV region) and thermal (in the meV region)
projectile
energies the competition between the different reaction mechanisms is quite
complicated. Normally the CN formation prevails below projectile energies of
approximately 10 to 20\,MeV. However, for light nuclei and magic nuclei the CN
formation is often suppressed because there are no CN levels that can be
populated. In this case the DI reaction mechanism
can dominate the nuclear reaction.

\section{Nuclear--structure information}
The nuclear structure of the compound nucleus $^{49}$Ca is of special interest
because of the expected simple single--particle character of the low--lying
states formed by coupling 2p$_{3/2}$, 1f$_{5/2}$ and 2p$_{1/2}$ neutrons to the
$J^\pi = 0^+$ ground state of $^{48}$Ca. At higher excitation energies many
bound levels can be described by 2particle--1hole configurations \cite{c}.
Information on neutron--unbound states in $^{49}$Ca ($S_{\rm n} = 5.142$\,MeV)
 comes from three different experimental sources: neutron--capture and
transmission measurements \cite{b,f,g,h}, (d,p)--reaction work \cite{c,i},
and high--resolution spectroscopy of $\beta$--delayed neutron ($\beta$dn)
decay \cite{d,e} of $J^\pi = 3/2^+$ in $^{49}$K. As is discussed in
Ref. \cite{l}, the $^{49}$K$(\beta^-)^{49}$Ca(n)$^{48}$Ca decay
can be considered as inverse process to s-- and d--wave neutron
capture on $^{48}$Ca. Companion theoretical studies to Ref. \cite{b} by
Divadeenam et al. \cite{m} (2p--1h doorway model), to Refs. \cite{d,e} by
Dobado
and Poves \cite{n} (complete sd--fp shell model),
as well as own QRPA shell--model
calculations (Gamow--Teller strength), using the code of M\"oller and Randrup
\cite{o}
agree quite well with the experimental observations.
The most remarkable result is
the absence of $J^\pi = 1/2^+$ states (s--wave resonances) up to at least
$S_{\rm n} + 800$\,keV, which can be understood in terms of the specific
quasiparticle
structure of $^{49}$Ca. The second observation is that the d--wave strength is
considerably larger than the p--wave strength. This is due to the existence of
a
d--wave giant resonance for $A = 48$. Finally, no strong neutron--resonances
were identified below 158\,keV. The two very small resonances at 20\,keV
and 107\,keV only found in neutron capture \cite{g}, but not in
$\beta$--delayed neutron decay \cite{d,e} and not in the (d,p)--reaction
\cite{c,i}, are probably also d--wave
resonances with $\Gamma_{\rm n} < \Gamma_\gamma$.

With this nuclear--structure information on $^{49}$Ca average continuum (HF)
and
resonance (Breit--Wigner, BW) neutron--capture rates were derived using the
code SMOKER \cite{l,p}. When comparing these values with the measured 30\,keV
Mawellian--averaged capture cross section of $^{48}$Ca (See Refs. \cite{f,g}),
one can draw the following conclusions: The statistical--model (HF) prediction
agrees with the measured cross section at 30 keV of about 1\,mb; this result
must, however, be
regarded as completely fortuitous. The contribution of the resonances, i.e. the
BW cross section, is only about 5\% of the total neutron--capture rate.
Hence, in the case of the doubly--magic nucleus
$^{48}$Ca, 95\% must be due to direct reaction processes.

In contrast to $^{48}$Ca(n,$\gamma$), the nuclear structure information on the
radioactive target $^{50}$Ca and the compound nucleus $^{51}$Ca are scarce.
Some
bound levels without spin assignment are known from reaction work and $^{51}$K
$\beta$--decay \cite{q}, and the singles spectrum of $\beta$dn's has been
measured (see Fig.\ \ref{fig1}, taken from Ref.\ \cite{d}; the continuum
underlying the peaks in the spectrum is due to the response of the spectrometer
used). Due to
Gamow-Teller selection rules, the peaks seen in the spectrum correspond to
$J^\pi = 1/2^+$ and $3/2^+$ neutron--unbound states in $^{51}$Ca. According to
the inverse relationship between $\beta$dn--decay and neutron capture \cite{d},
these states represent s-- and d--wave resonances in
$^{50}$Ca(n,$\gamma)^{51}$Ca.
Similar to the compound nucleus $^{49}$Ca, the most remarkable result is the
absence of $J^\pi = 1/2^+$ states up to 0.85\,MeV beyond the neutron separation
energy ($S_{\rm n} = 4.4$\,MeV \cite{r}). This results in a resonance
capture cross section of $\langle \sigma^{{\rm CN}} \rangle \simeq 8.5 \cdot
10^{-15}\,$mb when using the BW--formalism of the SMOKER code \cite{l,p}.
If this were the total capture rate for $^{50}$Ca, in any astrophysical
s-- and n$\beta$--type neutron--capture process \cite{f,j} the build--up
of $A > 50$ Ca isotopes would be strongly hindered by successful competition
of $\beta$--decay. With this, a strong overabundance of $^{50}$Ti --- as
observed in certain meteoritic inclusions
(for discussion see Refs. \cite{j,k}) --- would result.

However, also in the case of $^{50}$Ca neutron capture, it is not unlikely that
$\langle \sigma^{{\rm CN}} \rangle$ represents only a small fraction of the
total cross
section which may be dominated by the DI reaction rate. Shell--model
considerations (using the QRPA code of M\"oller et al. \cite{o,s})
support this possibility. When assuming the compound nucleus $^{51}$Ca
to be spherical, its ground state has the $\nu$p$_{3/2}$ configuration
($J^\pi = 3/2^-$). The two lowest excited levels are predicted
to be the $\nu$p$_{1/2}$ and $\nu$f$_{5/2}$ shell--model states at about
1.177\,MeV
and 1.209\,MeV, respectively. From the $0^+$ ground state of $^{50}$Ca indeed
strong p--wave radiative capture can be expected to the $J^\pi = 3/2^-$ and
$1/2^-$ levels in $^{51}$Ca. Another interesting result of our QRPA
calculations is the prediction of the lowest--energy $J^\pi = 1/2^+$ and
$3/2^+$ states of $\nu$g$_{9/2}$ shell--model origin around 5.50\,MeV
in $^{51}$Ca. This is in good agreement with our experimental result from the
$^{51}$K $\beta$dn--spectrum (Fig.\ \ref{fig1}).
After correction for recoil, we obtain for the two lowest neutron--unbound
$J^\pi = 1/2^+$, $3/2^+$ states $E_1^* = (S_{\rm n} + E_{{\rm n}1})
\simeq 5.31\,$MeV and $E_2^* = (S_{\rm n} + E_{{\rm n}2}) \simeq 5.37\,$MeV.
The neutron peaks at 1.18\,MeV,
1.46\,MeV, 2.20\,MeV and 2.48\,MeV identified in our $\beta$dn--spectrum
correspond to transitions to the $2_1^+$--state in $^{50}$Ca and,
hence, are not important for the
present resonance neutron--capture considerations.

In the following, we will present our direct neutron--capture calculations
using as input parameters the experimental and theoretical
information given above.

\section{DI-Calculations}
Potential models have often been used to describe direct reactions at
thermonuclear and
thermal projectile energies (Ref.\ \cite{t} and references therein).
They are based on the description of the dynamics of the reaction by
a Schr\"odinger equa\-tion with optical
potentials in the entrance and exit channel. Such models are the Distorted Wave
Born Approximation (DWBA) for transfer reactions or the Direct Capture Model
(DC) for radiative capture.

The DC cross section ist given by \cite{u}
\begin{eqnarray}
\sigma^{{\rm DC}} & = & \int {\rm d}\Omega \,
\frac{{\rm d}\sigma^{{\rm DC}}}{{\rm d}
\Omega_\gamma} \nonumber \\
& = & \int {\rm d}\Omega\,2\,\left( \frac{e^2}
{\hbar\,{\rm c}} \right) \left( \frac{\mu {\rm c}^2}{\hbar\,{\rm c}} \right)
\left( \frac{{\rm k}_\gamma}{{\rm k}_a} \right)^3 \frac{1}{2\,I_A + 1}\,
\frac{1}{2\,S_a + 1} \sum_{M_A\,M_a\,M_B,\,\sigma}
\mid T_{M_A\,M_a\,M_B,\,\sigma} \mid^2. \label{a}
\end{eqnarray}
The quantities $I_A,\;I_B$ and $S_A$ ($M_A,\;M_B$ and $M_a$) are the spins
(magnetic quantum numbers) of the target nucleus $A$, residual nucleus $B$
and projectile $a$, respectively. The reduced mass in the entrance channel
is given by $\mu$. The polarisation $\sigma$ of the electromagnetic radiation
can be $\pm 1$. The wave number in the entrance channel and for the emitted
radiation is given by $k_a$ and $k_\gamma$, respectively.

The multipole expansion of the transition matrices $T_{M_A\,M_a\,M_B,\,\sigma}$
including electric dipole (E1) and quadrupole (E2) transitions as well as
magnetic dipole (M1) transitions is given by
\begin{equation}
T_{M_A\,M_a\,M_B,\,\sigma} = T^{{\rm
E}1}_{M_A\,M_a\,M_B,\,\sigma}\,d^1_{\delta\,\sigma}
(\theta) + T^{{\rm E}2}_{M_A\,M_a\,M_B,\,\sigma}
d^2_{\delta\,\sigma}(\theta) + T^{{\rm M}1}_{M_A\,M_a\,M_B,\,\sigma}
d^1_{\delta\,\sigma}(\theta).
\end{equation}
The rotation matrices depend on the angle between $\vec{k}_a$ and
$\vec{k}_\gamma$
which is denoted by $\vartheta$, where $\delta = M_A + M_a - M_B$.

Defining
\begin{eqnarray}
C({\rm E}1) = i\,\mu\,\left( \frac{Z_a}{m_a} - \frac{Z_A}{m_A} \right),
C({\rm E}2) = \frac{{\rm k}_\gamma}{\sqrt{12}}\,\mu^2\,\left( \frac{Z_a}
{m_a^2} + \frac{Z_A}{m_A^2} \right),
\end{eqnarray}
we can write for the transition matrices for the electric dipole
(E${\cal L}$ = E1) or quadrupole (E${\cal L}$ = E2) transition
\begin{eqnarray}
T^{{\rm E}\cal L}_{M_A\,M_a\,M_B,\,\sigma} & = & \sum_{l_a\,j_a}
i^{l_a} \langle l_a\,0\,S_a\,M_a \mid j_a\,M_a \rangle
\langle j_b\,M_B - M_A\,I_A\,M_A \mid I_B\,M_B \rangle\nonumber\\
& &\nonumber\\
& \times & \langle {\cal L}\,\delta\,j_b\,M_B - M_A \mid j_a\,M_a \rangle
C(E{\cal L})\,\hat l_a\,\hat l_b\,\hat j_b\nonumber\\
& &\nonumber\\
& \times & \langle l_b\,0\,{\cal L}\,0 \mid l_a\,0 \rangle W({\cal
L}\,l_b\,j_a\,S_a;l_a\,j_b)
I^{E\cal L}_{l_b\,j_b\,I_B;l_a\,j_a}. \label{aa}
\end{eqnarray}
In the above expressions $Z_a,\;Z_A$ and $m_a,\;m_A$ are the charge and
mass numbers of
the projectile $a$ and target nucleus $A$, respectively. The quantum numbers
for the channel spin in
the entrance channel and for the transferred angular momentum are denoted by
$j_a$ and $j_b$, respectively. Here and in the following the
abbreviation $\hat l$ stands for $\sqrt{2l+1}$ (the same applies to other
quantum numbers).

For magnetic dipole transitions we obtain
\begin{eqnarray}
T^{{\rm M}1}_{M_A\,M_a\,M_B,\,\sigma} & = & \sum_{l_a\,j_a}\,i^{l_a}\,\sigma
\Biggl[ \langle l_a\,0\,S_a\,M_a \mid j_a\,M_a \rangle \langle j_b\,M_B -
M_A\,I_A\,M_A
\mid I_B\,M_B \rangle \Biggr.\nonumber\\
& &\nonumber\\
& \times & \langle 1\,\delta\,j_b\,M_B - M_A \mid j_a\,M_a \rangle\nonumber\\
& &\nonumber\\
& \times & \Biggl\{ \mu\left( \frac{Z_A}{m_A^2} + \frac{Z_a}
{m_a^2} \right)\,\hat l_b\,\hat j_b\,\sqrt{l_a\,(l_a + 1)}
W(1\,l_a\,j_a\,S_a;l_a\,j_b) \nonumber\\
& &\nonumber\\
& + & 2\mu_a\,(-1)^{j_b - j_a}\,\hat S_a\,\hat{j_b}\,\sqrt{S_a\,(S_a + 1)}
W(1\,S_a\,j_a\,l_a;S_a\,j_b) \Biggr\}\nonumber\\
& &\nonumber\\
& - & \langle l_a\,0\,S_a\,M_a \mid j_a\,M_a \rangle
\langle j_a\,M_a\,I_A\,M_B - M_a
\mid I_B\,M_B \rangle\nonumber\\
& &\nonumber\\
& \times & \langle I_A\,M_B - M_a\,1\,\delta \mid I_A\,M_A \rangle \nonumber\\
& &\nonumber\\
& \times & \left. \mu_A\,\delta_{j_a\,j_b}\,\sqrt{\frac{I_A + 1}{I_A}} \right]
\left[ \frac{\hbar\,{\rm c}}{2\,m_{\rm p}\,{\rm c}^2}
\right] \delta_{l_a\,l_b}\,
\hat l_a\,I^{M1}_{l_b\,j_b\,I_B;l_a\,j_a}, \label{ab}
\end{eqnarray}
where $W$ is the Racah coefficient, the $\mu_i$ are the magnetic moments and
$m_{\rm p}$
is the mass of the proton.

The overlap integrals in Eqs.\ (\ref{aa},\ref{ab}) are given by \cite{v}
\begin{equation}
I^{{\rm E}{\cal L}}_{l_b\,j_b\,I_B;l_a\,j_a} =
\int\,{\rm d}r\,U_{l_b\,j_b\,I_B}(r)
{\cal O}^{{\rm E}{\cal L}}(r)\,\chi_{l_a\,j_a}(r) \label{ac}
\end{equation}
for the electric dipole (E${\cal L}$ = E1) or quadrupole (E${\cal L}$ = E2)
transition,
and by
\begin{equation}
I^{{\rm M}1}_{l_b\,j_b\,I_B;l_a\,j_a} = \int\,{\rm d}r\,U_{l_b\,j_b\,I_B}(r)
{\cal O}^{{\rm m}1}(r)\,\chi_{l_a\,j_a}(r)
\end{equation}
for the magnetic dipole transition (M${\cal L}$ = M1).
The radial part of the bound state wave function in the exit channel and the
scattering
wave function in the entrance channel is given by $U_{l_b\,j_b\,I_B}(r)$ and
$\chi_{l_a\,j_a}(r)$, respectively. The radial parts of the electromagnetic
multipole operators are
\begin{eqnarray}
{\cal O}^{{\rm M}1}(r) & = & \frac{1}{2\rho}\,\left[ \sin\,\rho +
\rho\,\cos\,\rho \right],\\
{\cal O}^{{\rm E}1}(r) & = & \frac{3}{\rho^3}\,\left[ (\rho^2 - 2)\,\sin\,\rho
+ 2\,\rho\,\cos\,\rho \right] r,\\
{\cal O}^{{\rm E}2}(r) & = & \frac{15}{\rho^5}\,\left[ (5\,\rho^5 - 12)\,\sin\,
\rho + (12 - \rho^2)\,\rho\,\cos\,\rho \right] r^2.
\end{eqnarray}
In the long wavelength approximation --- applicable in our case, since $\rho =
k_\gamma r \ll 1$ --- these quantities reduce to
\begin{eqnarray}
{\cal O}^{{\rm M}1}(r) & \simeq & 1,\\
{\cal O}^{{\rm E}1}(r) & \simeq & r,\\
{\cal O}^{{\rm E}2}(r) & \simeq & r^2.
\end{eqnarray}
The most important ingredients in the potential models are the
wave functions for the scattering and bound states in the entrance and exit
channels. In calculations performed by our group the potentials are determined
by using the folding procedure
\cite{t}. In this approach the number of open parameters is reduced
considerably
compared to more phenomenological potentials (e.g. Saxon--Woods potentials).
The
nuclear densities are derived from nuclear charge distributions \cite{w} and
folded with an energy and density dependent nucleon--nucleon (NN) interaction
$v_{\rm eff}$ \cite{x}:
\begin{equation}
V(R) = \lambda\,V_{\rm F}(R) = \lambda\,\int\int
\rho_a(\vec{r}_1)\rho_A(\vec{r}_2)\,
v_{\rm eff}\,(E,\rho_a,\rho_A,s)\,{\rm d}\vec{r}_1{\rm d}\vec{r}_2. \label{ad}
\end{equation}
The variable $s$ in the NN interaction term is given by
\begin{equation}
s = |\vec{R} + \vec{r}_2 - \vec{r}_1| \label{ae}
\end{equation}
with $\vec{R}$ being the separation of the centers of mass of the two colliding
nuclei.
The normalization factor $\lambda$ accounts empirically
for Pauli repulsion effects and dispersive
parts in the potential $V(R)$ that are not included in the folding potential
$V_{\rm F}(R)$. This parameter can be adjusted to elastic scattering data
and/or
bound and resonant state energies of nuclear cluster states and at the same
time
ensures the correct behaviour of the wave functions in the nuclear exterior.

For the calculation of the DI capture cross section
(Eqs.\ (\ref{a}--\ref{ad})) we used the code
TEDCA \cite{y}. The folding potentials (Eq.\ (\ref{ad})) were
determined with the help of the code VOLD \cite{z}.

The reaction $^{48}$Ca(n,$\gamma)^{49}$Ca was calculated for projectile
energies
below 250\,keV. For the potential $^{48}$Ca + n in the entrance and
exit channel we used
\begin{equation}
V(R) = \lambda\,V_{\rm F}(R) \label{af}
\end{equation}
where $V_{\rm F}$ is the folding potential of Eq.\ (\ref{ad}). The strength
of the potential $\lambda_{\rm i}$ in the entrance channel was determined using
neutron scattering data obtained from
\cite{h}. The data was fitted to pure s-wave scattering with $\sigma =
0.019$\,b resulting in a $\lambda_{\rm i}$
of 0.93357 and giving a volume integral of
436.9\,MeV\,fm$^3$. \\
For the optical potential in the entrance channel we neglected the
imaginary part potential because the flux into the other channels is very
small.
One may wonder, if the capture process would not itself produce a large
damping, because at low energies the capture cross section is larger
than the elastic cross section. For instance, at thermal energies the
neutron--capture cross section of $^{48}$Ca is about 1\,b, whereas the elastic
neutron cross section is about 0.02\,b. However, this can be explained by the
different phase--space factors for the capture and elastic cross
sections. For the capture cross section near threshold one can write
$\sigma_{\rm c} \approx (4\pi\,R/k_a)(-{\rm Im}f_0/|f_0|^2)$, where
$f_0$ is the logarithmic derivative at a radial distance $R$. The elastic
cross section is about constant for projectiles near threshold, whereas
the capture cross section has the well known 1/$v$--behaviour.
{}From the thermal value of the capture cross section, using $R = 4.5$\,fm,
$-{\rm Im}f_0/|f_0|^2 \approx 6 \cdot 10^{-5}$ is obtained.
Since this expression can be regarded as a measure of
damping in the entrance channel, we see that the damping effects are rather
small. Therefore, at low energies the large value of the capture cross section
(1/$v$--behaviour) compared to the elastic cross section (only weakly
energy--dependent) is due to the phase--space factors and does not result in
additional damping effects.\\
For the exit channels in $^{49}$Ca we fitted $\lambda_{\rm f}$ to reproduce the
experimental neutron separation energies in $^{49}$Ca. We obtained
$\lambda_{\rm f} = 0.9648,\,0.8863,\,0.8417,\,1.6745,\,1.6014$ and 1.5885
for the 2p$_{3/2}$, 2p$_{1/2}$, 1f$_{5/2}$, 2f$_{5/2}$, 3p$_{3/2}$ and
3p$_{1/2}$ state in $^{49}$Ca, respectively.
Spins, excitation energies, Q--values and spectroscopic factors for the
transitions to the ground and excited states
are listed in Table \ref{tab1} \cite{i}.

\section{Results and discussion}
The theoretical cross section $\sigma^{\rm th}$ is obtained from the direct
capture cross section $\sigma^{\rm DC}$ in Eq.\ (\ref{a}) as a sum over
each final state $i$ by
\begin{equation}
\sigma^{\rm th} = \sum_i C_i^2 S_i \sigma_i^{\rm DC}.
\end{equation}
In our case the isospin Clebsch-Gordan coefficients $C_i$ are 1.
The spectroscopic factors $S_i$ describe the overlap between the
antisymmetrized wave functions of $^{48}$Ca + n and the final state
in $^{49}$Ca. We obtained the values for the spectroscopic factors
given in Table \ref{tab1} from a recent
analysis of $^{48}$Ca(d,p)$^{49}$Ca \cite{i}.

The most important contributions to the direct capture
cross section are given by the
transitions to the ground and first excited state in $^{49}$Ca with a
Q--value of 5.142\,MeV and 3.121\,MeV, respectively (Table \ref{tab2}).
For the transition to the ground state (g.s.) we obtained the
Maxwellian--averaged direct capture cross section at 30\,keV
\begin{displaymath}
\langle \sigma_{\rm g.s.}^{\rm th} \rangle = 0.791\,{\rm mb},
\end{displaymath}
and for the transition to the first excited state at 2.021\,MeV
\begin{displaymath}
\langle \sigma_{\rm exc.}^{\rm th} \rangle = 0.222\,{\rm mb}.
\end{displaymath}
With the small contributions of the 3p$_{3/2}$ state at 4.069\,MeV and the
3p$_{1/2}$ state
at 4.261\,MeV we obtain
$\langle \sigma_{\rm exc.}^{\rm th} \rangle = 1.04\,{\rm mb}$ for the
Maxwellian--averaged direct capture cross section at 30\,keV.
The cross section for the
transitions to the other high--spin states in the final nucleus can be
neglected
(Table \ref{tab2}). This is due to the fact that only for final p--states an
E1--transition
with s--wave neutrons in the entrance channel is possible.

Fig.\ \ref{fig2}
shows the contributions to the direct capture at a projectile energy
of 30\,keV as a function of the radial distance from the target nucleus
given by the integrand of Eq.\ (\ref{ac}).
It is interesting to note that the important contributions to the investigated
capture reaction come from the nuclear surface (4.5--5\,fm) as well as from the
nuclear exterior.

In Table \ref{tab3} the theoretical DC cross sections $\sigma^{\rm th}$
obtained in this work
are compared to the experimental data $\sigma^{\rm exp}$. As can be seen from
Table \ref{tab3} and Fig.\ \ref{fig3} the experimental data can be reproduced
excellently for the thermal as well as for the thermonuclear energy region
by our DC--calculations. In Fig.\ \ref{fig3} the theoretical
direct capture cross section (solid line) is shown together with the
experimental data of Cranston and White \cite{a}, K\"appeler et al. \cite{f}
and Carlton et al. \cite{g}.
In this plot we find the well known 1/$v$ behaviour of the experimental as
well as the theoretical cross section ranging from the meV to the MeV region.

We also calculated the direct capture cross section for the reaction
$^{50}$Ca(n,$\gamma)^{51}$Ca.
To obtain the $^{50}$Ca + n--potential we used the density distribution
obtained from QRPA calculations \cite{o}.
The Q--value for the p$_{3/2}$ transition
to the ground state is given by 4.4\,MeV \cite{r}. The resonant states
$J^\pi = 1/2^+$ and $3/2^+$ of $\nu g_{9/2}$ shell--model origin around
5.50\,MeV mentioned in the introduction are of 2p--1h type and therefore cannot
be reached by direct capture. They have not been included in our calculation.
The spin and the excitation energy
of the first excited state is 1/2$^-$ at 1.1768\,MeV. The above values  and the
spectroscopic
amplitudes were obtained from
QRPA shell model calculations using a folded--Yukawa single--particle and a
Lipkin--Nogami pairing model \cite{o}. These input parameters
(see Table \ref{tab4}) yield
\begin{displaymath}
\langle \sigma^{\rm DC} \rangle = 0.7\,{\rm mb}
\end{displaymath}
for the Maxwellian--averaged cross section of $^{50}$Ca(n,$\gamma)^{51}$Ca at
30\,keV.
The lower value of this cross section compared to the cross section of 1.04\,mb
for $^{48}$Ca(n,$\gamma)^{49}$Ca is mainly due to the lower Q--value.

These results are important for astrophysical r-- and $\alpha$--process
calculations. Previously, the rates were calculated in the statistical
model which yields a vanishing cross section for
$^{50}$Ca(n,$\gamma)^{51}$Ca. Therefore, the turning point in the
neutron--capture path was at $^{48}$Ca because of the very small cross
sections of the heavier isotopes. With the above considerably higher cross
section (comparable to the one of $^{48}$Ca(n,$\gamma)^{49}$Ca) the path can
continue beyond $^{48}$Ca or even beyond $^{50}$Ca. Recent network
calculations prove that possibility at least for low entropies of the
Type II supernova hot entropy bubble~\cite{tommy}. This underlines the
importance of direct capture in astrophysical environments.

\section{Summary}
We used the reaction $^{48}$Ca(n,$\gamma)^{49}$Ca as a test case for
calculating the DC cross section, since for this reaction all the relevant
information is available
from various experiments. The experimental cross section could be reproduced
assuming a direct mechanism and using the potential model and the folding
procedure.

As a more general conclusion, we have confirmed for the $^{48}$Ca region
that the
applicability of the statistical assumptions in the commonly used HF
calculations to derive neutron--capture cross sections breaks down near
magic shells. Rather, the cross section is dominated by one or a few
resonances, or --- as in our cases --- by direct radiative capture to
bound final states in the absence of low--lying CN resonances.

As we have shown, some experimental information on neutron--capture
resonances --- even
far from stability --- can be obtained from the decay model of $\beta$--delayed
neutron emission. With regard to spectroscopic factors of bound
states, they may be obtained
from (d,p)--reactions in inverse kinematics using cooled radioactive beams.

\section*{Acknowledgement}
We are indebted to F.-K. Thielemann for
making his code SMOKER available to us.
We want to thank the Fonds zur F\"orderung wissenschaftlicher Forschung in
\"Osterreich
(project S7307--AST) and the \"Osterreichische Nationalbank (project 5054)
for their support. TR acknowledges support by the Alexander von Humboldt
foundation and by an APART fellowship of the Austrian Academy of
Sciences.

\newpage

\begin{figure}
\caption{Beta--delayed neutron spectrum of $J^\pi = 1/2^+$ $^{51}$K decay.}
\label{fig1}
\end{figure}

\begin{figure}
\caption{Contributions to the direct--capture cross section at 30\,keV
projectile energy as a function of the radial distance from the target
nucleus.}
\label{fig2}
\end{figure}

\begin{figure}
\caption{Comparison of the calculated direct--capture cross section
(solid line) with the experimental data of $^{48}$Ca(n,$\gamma)^{49}$Ca
for projectile energies from the meV to the MeV region [1],[6],[7].}
\label{fig3}
\end{figure}

\begin{table}
\caption{Spins, excitation energies, Q--values and spectroscopic factors
of levels in $^{49}$Ca [9].}
\begin{tabular}{rddd}
$J^\pi$ & $E_{\rm ex}$ [MeV] & $Q$ [MeV] & $S$ \\ \hline
2p$_{3/2}$ & 0.000 & 5.142 & 0.84 \\
2p$_{1/2}$ & 2.021 & 3.121 & 0.91 \\
1f$_{5/2}$ & 3.586 & 1.556 & 0.11 \\
2f$_{5/2}$ & 3.993 & 1.149 & 0.84 \\
3p$_{3/2}$ & 4.069 & 1.073 & 0.13 \\
3p$_{1/2}$ & 4.261 & 0.881 & 0.12 \\
\end{tabular}
\label{tab1}
\end{table}

\begin{table}
\caption{Maxwellian--averaged direct--capture cross sections for neutron
capture in $^{48}$Ca.}
\begin{tabular}{rdd}
$J^\pi$ & $\langle \sigma(25\,{\rm keV}) \rangle$ [mb] &
$\langle \sigma(30\,{\rm keV})
\rangle$ [mb] \\ \hline
2p$_{3/2}$ & 0.872 & 0.791 \\
2p$_{1/2}$ & 0.244 & 0.222 \\
1f$_{5/2}$ & 0.000 & 0.000 \\
2f$_{5/2}$ & 0.000 & 0.000 \\
3p$_{3/2}$ & 0.022 & 0.020 \\
3p$_{1/2}$ & 0.007 & 0.006 \\ \hline
Sum & 1.145 & 1.039 \\
\end{tabular}
\label{tab2}
\end{table}

\begin{table}
\caption{Theoretical and experimental cross section for the capture reaction
$^{48}$Ca(n,$\gamma)^{49}$Ca.}
\begin{tabular}{ccdd}
ENERGY$^{\rm a}$ & REFERENCE OF & EXPERIMENTAL & DIRECT CAPTURE \\
& EXPERIMENTAL DATA & CROSS SECTION & CROSS SECTION \\ \hline
0.0253\,eV (M. a.) & Cranston and White \cite{a} & (1.09 $\pm$ 0.14)\,b &
1.13\,b \\
25\,keV (M. a.) & K\"appeler et al. \cite{f} & (1.03 $\pm$ 0.09)\,mb &
1.15\,mb \\
30\,keV (M. a.) & Carlton et al. \cite{g} & (1.05 $\pm$ 0.13)\,mb & 1.04\,mb \\
97\,keV & K\"appeler et al. \cite{f} & (0.55 $\pm$ 0.09)\,mb & 0.58\,mb \\
\end{tabular}
\label{tab3}
\tablenotetext[1]{M. a. Maxwellian averaged.}
\end{table}

\begin{table}
\caption{Spins, excitation energies, Q--values and spectroscopic factors of
levels in $^{51}$Ca
[15].}
\begin{tabular}{rddd}
$J^\pi$ & $E_{\rm ex}$ [MeV] & $Q$ [MeV] & $S$ \\ \hline
2p$_{3/2}$ & 0.000 & 4.400 & 0.80 \\
2p$_{1/2}$ & 1.177 & 3.223 & 0.92 \\
\end{tabular}
\label{tab4}
\end{table}

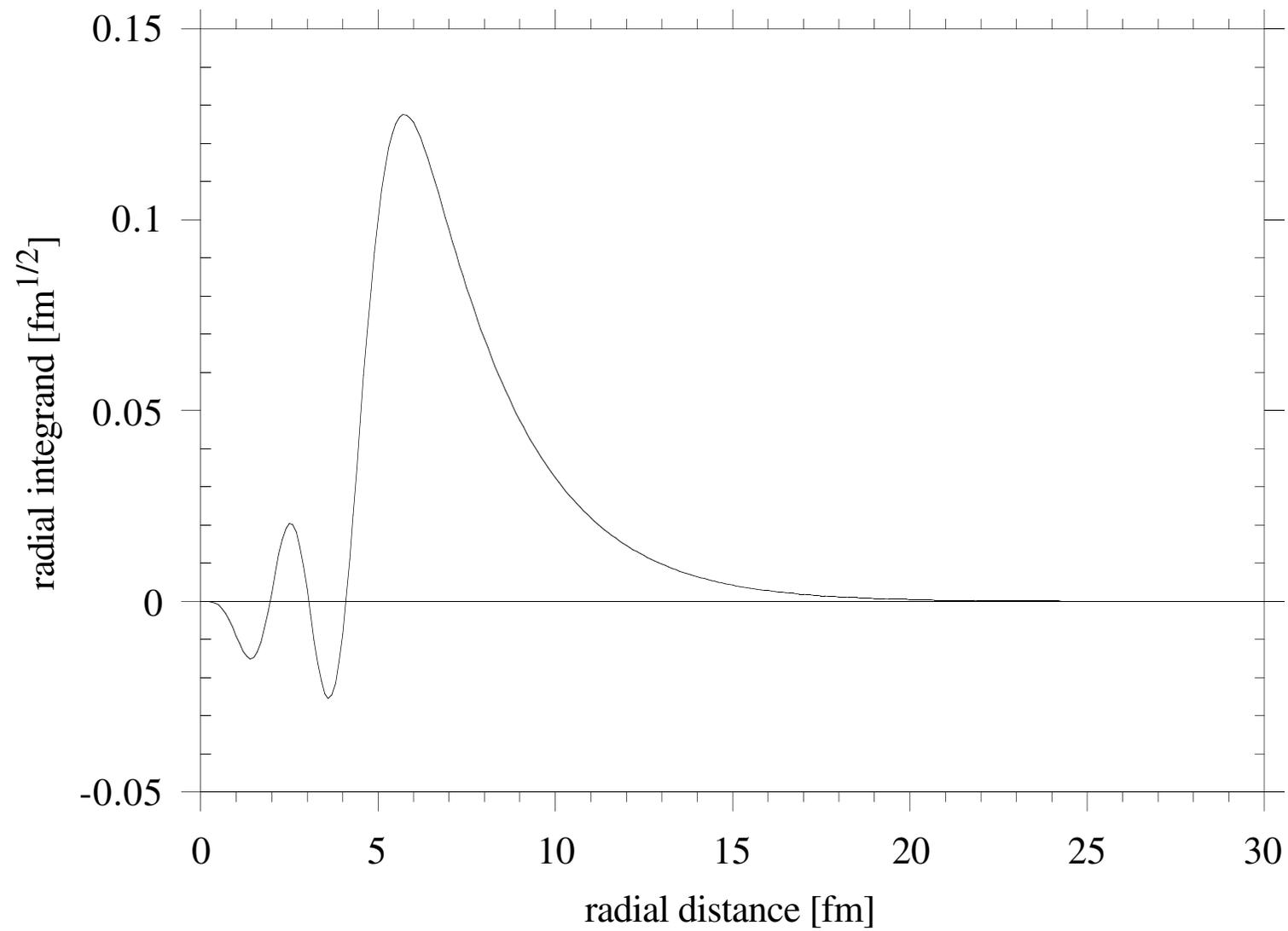
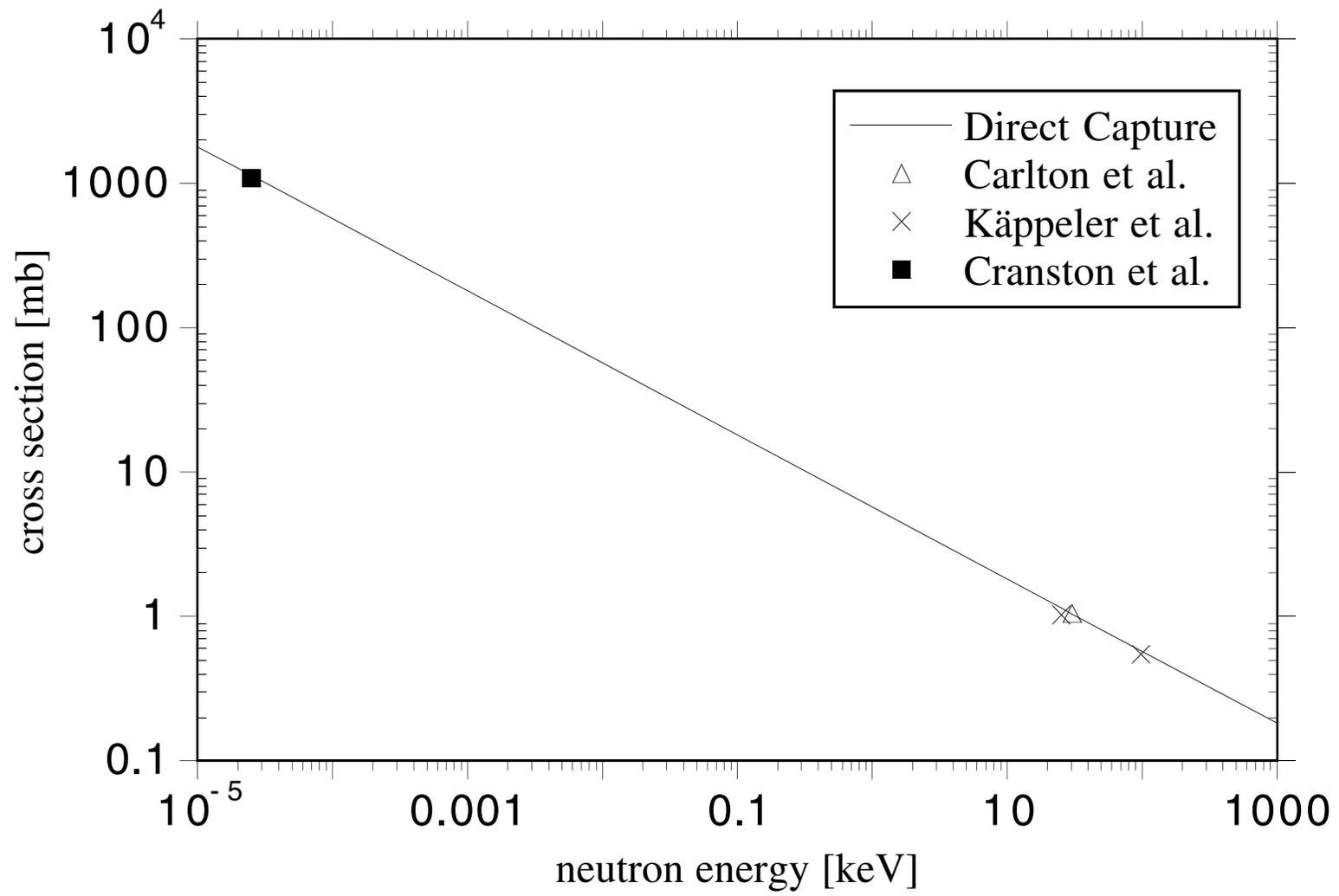
\end{document}